# SLM aided noninvasive imaging through thin scattering layers

**SASWATA MUKHERJEE**[*,1,2], **A. VIJAYAKUMAR**[1,2], **AND JOSEPH ROSEN**[1]

[1]School of Electrical and Computer Engineering, Ben-Gurion University of the Negev, P.O. Box 653, Beer-Sheva 8410501, Israel.
[2]The authors contributed equally to the work.
[*saswata7@gmail.com](mailto:saswata7@gmail.com)

**Abstract:** We propose and demonstrate a new imaging technique to noninvasively see through scattering layers with the aid of a spatial light modulator (SLM). A relay system projects the incoherent light pattern emitting from the scattering layer onto the SLM. Two coded phase masks are displayed, one after another, on the SLM to modulate the projected scattered field. Two corresponding intensity patterns are recorded by a digital camera, and subtracted one from the other in the computer to obtain a bipolar matrix. A modified phase retrieval algorithm is used to retrieve the object information from this bipolar matrix.

## 1. Introduction

Imaging through scattering layers enables seeing objects present behind media such as biological tissues [1-3], fog [4], and other turbid media [5]. Different imaging techniques have been developed in the past to see through scattering layers which can be broadly classified into invasive and noninvasive categories depending upon whether the point spread function (PSF) of the system is measured or not. In the case of invasive techniques, it is necessary to have prior information about either the PSF of the imaging system containing the scattering layer or the phase image of the scattering layer [6-11]. Besides being invasive, these techniques are not suitable for imaging through temporally varying turbid media, like fog, or a blood flow, whose PSF varies with time. Furthermore, a high level of reconstruction noise was noticed [12,13] in some of the recently developed 2D and 3D imaging techniques using cross-correlations between the PSF and object intensity response. Consequently, different techniques were developed to suppress the reconstruction noise by engineering the phase masks [14], and statistical averaging [15], which are not always practical in the case of impenetrable scattering layers. Other deconvolution techniques based on Richardson-Lucy iterative algorithm [16] and inverse filter [17] are invasive as well, as they require to measure the PSF in order to retrieve the image of the object.

Noninvasive imaging techniques have been developed recently to see through scatterers without knowing the PSF of the system or the phase function of the scattering layers [18-26]. In some of these noninvasive techniques, the intensity of the scattered light is autocorrelated and from this autocorrelation, the image of the concealed object is reconstructed using a phase retrieval algorithm [27,28]. While the noninvasive technique can retrieve the image without knowing the PSF, the success of the phase retrieval algorithm is highly statistical as the outcome of the phase retrieval method is dependent upon the initial guess of the iterated matrix. Consequently, the phase retrieval algorithm must run several times with different initial random patterns before the optimal result can be obtained [21]. One of the main requirements for the phase retrieval based noninvasive imaging techniques to be successful is that the phase variation of the scattering layers must be chaotic such that the autocorrelation of the system PSF yields a delta-like function. This requirement highly limits the applicability of the technique. In a formal notation, for incoherent illumination and for object intensity $O$, the intensity distribution recorded by the camera is $O*I_{PSF}$, where $I_{PSF}$ is the system PSF and '$*$' is the sign of 2D convolution. In [20,21], it is suggested to compute the autocorrelation

$(O*I_{PSF}) \otimes (O*I_{PSF})$, where '$\otimes$' represents the 2D correlation operator. The same autocorrelation can be written as $(O \otimes O)*(I_{PSF} \otimes I_{PSF})$ [29]. Under the assumption that $(I_{PSF} \otimes I_{PSF})$ is equal to a delta-like function ($\delta$), the autocorrelation of the camera output is approximately $(O \otimes O)$ and hence the object distribution can be retrieved by the phase retrieval algorithm. However, the assumption that $(I_{PSF} \otimes I_{PSF}) \approx \delta$ is sometimes problematic. First, the autocorrelation of a positive chaotic PSF has non-zero background distribution with double the size of $I_{PSF}$. Second, the width of the delta-like function is dependent on the properties of the scattering layer and hence the approximation to a narrow delta-like function is limited only to certain scatterers. Because of these two effects, the approximation $(I_{PSF} \otimes I_{PSF}) \approx \delta$ is sometimes not established well, and hence the input of the phase retrieval algorithm is far from the ideal $O \otimes O$. Consequently, the output of the algorithm might not converge to the required image of $O$.

In this study, we present a new approach to extend the applicability of the phase retrieval based noninvasive imaging techniques for imaging through various types of scattering layers. The main concept of the proposed method is to project the scattered light emitted from the scattering layer on a computer-controlled coded phase mask (CPM) which is used as a second scattering mask. By doing that, we intend to achieve two goals. First, since the second mask is synthesized digitally, we have better control on the properties of the combined scattering medium, such that a better approximation to the delta function can be obtained from the autocorrelation of the PSF. Second, in our proposed procedure, two camera shots, with two different CPMs, are captured and subtracted one from the other. Hence, the obtained superposed distribution $O*[I_{PSF1}-I_{PSF2}]$ is bi-polar with negligible bias term. Consequently, the background of the autocorrelation of the PSF is much lower than the background obtained from the autocorrelation of a positive function. These two improvements in the original method bring the autocorrelation of the system response closer to the autocorrelation of the object. Thus, the proposed method increases the chances of the phase retrieval algorithm to converge to an image, which is more similar to the original object.

## 2. Methodology and analysis

The optical configuration of the proposed setup is shown in Fig. 1. A spatially incoherent light [30] critically illuminates an object using a lens $L_1$. The light diffracted from the object is incident on a diffuser used as the scattering layer and located at a distance of $z_s$ from the object. A relay system projects the light pattern obtained at a distance of $z_r$ from the diffuser onto a phase-only spatial light modulator (SLM). The relay system can either be constructed from two identical lenses $L_2$ and $L_3$ with an overall magnification of 1, like in our experiment or if required, from different lenses with a magnification different from 1. The phase mask displayed on the SLM is a modulo-$2\pi$ phase addition of the CPM, synthesized using modified Gerchberg-Saxton algorithm (GSA) [31], and a quadratic phase mask (QPM) with the focal distance $f$ used to focus the light on an image sensor. The light modulated by the SLM is incident on the image sensor located at a distance of $z_h$ from the SLM.

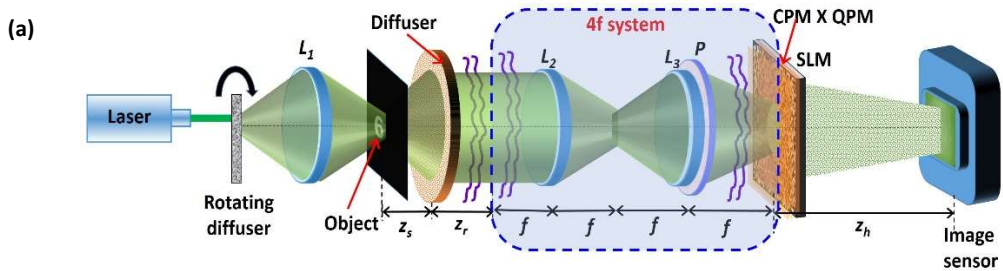

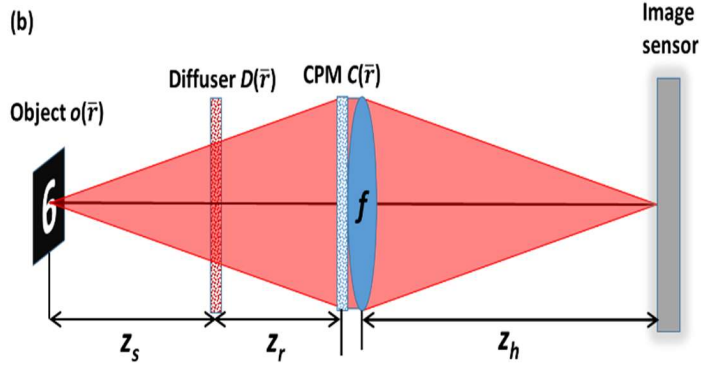

Fig. 1. (a) Optical configuration of the imaging setup. (b) Simplified optical scheme of (a) used for the analysis of the setup; CPM – Coded phase mask; $L_1$, $L_2$, $L_3$ – Refractive lenses; P – Polarizer; SLM – Spatial light modulator and QPM – Quadratic phase mask.

*2.1 System analysis*

The goal of the following mathematical formalism is to obtain the expression of the intensity distribution over the sensor plane for any arbitrary object and diffuser functions. Knowing this distribution might improve our understanding of the parameters which contribute to the quality of the reconstructed images. A simplified scheme for calculating the intensity over the sensor plane is given in Fig. 1(b), in which the QPM is symbolized by a lens with a focal length of $f$. Furthermore, the illuminating sub-system and the relay are omitted since these components have no influence on the mathematical analysis. We choose to represent the object intensity distribution $o(\bar{r})$ as a series of delta functions, whereas the diffuser $D(\bar{r})$ and the CPM $C(\bar{r})$ are represented as Fourier series of linear phases, as following,

$$o(\bar{r})=\sum_{j=1}^{N}a_j\delta(\bar{r}-\bar{r}_j),\ D(\bar{r})=\sum_{j=-\infty}^{\infty}b_j\exp(i2\pi\bar{v}_j\cdot\bar{r}),\ C(\bar{r})=\sum_{j=-\infty}^{\infty}c_j\exp(i2\pi\bar{v}_j\cdot\bar{r}). \qquad(1)$$

Assume the imaging equation $(1/u)+(1/z_h)=1/f$ is fulfilled between the object and the sensor planes, where $u=z_s+z_r$ and the distances $z_s$, $z_r$ and $z_h$ are defined in Fig. 1(b). For a single point source at $\bar{r}_j$ and a single linear phase at each of the two phase masks $D(\bar{r})$ and $C(\bar{r})$, the image on the sensor is the image of the source point but shifted according to the parameters of the linear phases. This shift can be expressed by the convolution operator as the following,

$$I(\bar{r}_o)=a_j\delta\left(\bar{r}_o-\frac{z_h\bar{r}_j}{u}\right)*\left|b_k\delta\left(\bar{r}_o-\frac{\lambda z_h z_s \bar{v}_k}{u}\right)*c_l\delta(\bar{r}_o-\lambda z_h \bar{v}_l)\right|^2, \qquad(2)$$

where $\lambda$ is the central wavelength of the illumination light. For the entire points of the object and for the entire linear phases composing the phase masks, the intensity on the camera is a convolution between the three series, as follows,

$$I(\bar{r}_o)=\sum_{j}^{N}a_j\delta\left(\bar{r}_o-\frac{z_h\bar{r}_j}{u}\right)*\left|\sum_{j}^{\infty}b_j\delta\left(\bar{r}_o-\frac{\lambda z_h z_s \bar{v}_j}{u}\right)*\sum_{j}^{\infty}c_j\delta(\bar{r}_o-\lambda z_h \bar{v}_j)\right|^2. \qquad(3)$$

The most left series represent the object according to Eq. (1), whereas the next two series represent the Fourier transforms of the diffuser with the scaling operator of $v[u/\lambda z_s z_h]$, and of the CPM with the scaling operator of $v[1/\lambda z_h]$. The scaling operator is defined by the equation $v[\alpha]f(x)=f(\alpha x)$. Therefore, the intensity on the camera is

$$I(\bar{r}_o) = o(\bar{r}_o) * \left| v\left[\frac{u}{\lambda z_h z_s}\right] \mathfrak{F}\{D(\bar{r})\} * v\left[\frac{1}{\lambda z_h}\right] \mathfrak{F}\{C(\bar{r})\} \right|^2$$
$$= A_o o(\bar{r}_o) * \left| \mathfrak{F}\left\{ D\left(\frac{\lambda z_h z_s \bar{r}}{u}\right) \cdot C(\lambda z_h \bar{r}) \right\} \right|^2, \quad (4)$$

where $\mathfrak{F}$ is the 2D Fourier transform operator and $A_o$ is a constant. Based on Eq. (4), the intensity response of the system to a point source in the origin known as the PSF is,

$$I_{PSF}(\bar{r}_o) = \left| \mathfrak{F}\left\{ D\left(\frac{\lambda z_h z_s \bar{r}}{u}\right) \cdot C(\lambda z_h \bar{r}) \right\} \right|^2. \quad (5)$$

*2.2 Engineering of the autocorrelation profile using CPM*

Equation (5) indicates that the intensity response obtained by the given diffuser can be modified by introducing a CPM into the system. As explained in the introduction, the reconstruction algorithm is efficient if the autocorrelation of $I_{PSF}$ is as close as possible to a delta function. Hence, the CPM is introduced into the system in order to bring the autocorrelation of $I_{PSF}$ as close as possible to a delta function. Recall that the autocorrelation of a positive random function includes a background distribution around the delta-like response, we suggest acquiring two camera shots with two independent CPMs and to subtract one response from the other. By doing that, one can get a bipolar PSF with a negligible bias, in which its autocorrelation is lack of the undesired background distribution. Therefore, two intensity patterns are recorded, and one is subtracted from the other to obtain a bipolar intensity pattern $H_{OBJ}$ given by,

$$H_{OBJ}(\bar{r}_0) = I_{OBJ,1}(\bar{r}_0) - I_{OBJ,2}(\bar{r}_0)$$
$$= o(\bar{r}_o) * \left[ \left| \mathfrak{F}\left\{ D\left(\frac{\lambda z_h z_s \bar{r}}{u}\right) \cdot C_1(\lambda z_h \bar{r}) \right\} \right|^2 - \left| \mathfrak{F}\left\{ D\left(\frac{\lambda z_h z_s \bar{r}}{u}\right) \cdot C_2(\lambda z_h \bar{r}) \right\} \right|^2 \right], \quad (6)$$

where $C_i(\bar{r})$ is the phase function of the $i^{th}$ CPM.

The GSA algorithm shown schematically in Fig. 2 is used to synthesize the CPMs with different degrees of scattering. The GSA is implemented between the plane of the CPM and the sensor to obtain on the sensor plane an intensity pattern with a limited area. Since the sensor plane is the spectral domain of the CPM, the area limit imposed on the sensor plane controls the maximal scattering angle of the synthesized CPM. Assuming the width $B_x$ along $x$ is equal to the height $B_y$ along $y$, the degree of scattering σ is defined as the ratio $B/B_{max}$, where $B=B_x=B_y$ and $B_{max}$ is the maximum possible value of $B$. The phase images of the CPMs synthesized using GSA for σ=0.05, 0.1, 0.25, 0.5 and 1 are shown in Figs. 3(a)-3(e), respectively.

Apparently, the basic assumption of the phase-retrieval-like algorithms in which the autocorrelation of the PSF has a shape of a delta function is not always satisfied. We show the difficulties with this assumption by computing the PSF for the above mentioned five degrees of scattering, and for two possible setups, with and without the diffractive converging lens with the focal length $f$ (the QPM in Fig. 1(a) and the lens in Fig. 1(b)). The purpose of the QPM is to focus the modulated light onto the sensor and by that to increase the SNR of the imaging. The case without the QPM is considered herein because this is the configuration of several other experiments [20,22-25]. If the basic assumption is not valid, the convergence of the phase retrieval algorithm to the real hidden object becomes rare and when it happens an automatic algorithm cannot know if it is the true expected reconstruction. Next, we show by computer simulation and by laboratory experiments that by introducing CPM into the system, the autocorrelation of the new PSF becomes more similar to a delta-like function.

A point object is introduced into the simulated system, and its complex amplitude propagates through the different optical components of the optical configuration shown in Fig. 1. In both cases, with and without the QPM, the scattering medium is positioned at the SLM plane and this medium is tested for 5 different degrees of scattering. For both configurations, the intensity patterns ($I_{PSF}$) and their corresponding autocorrelation profiles $(I_{PSF} \otimes I_{PSF})$ are depicted for different degrees of scattering, as shown in Fig. 4 and Fig. 5. The plots of the cross-section of $(I_{PSF} \otimes I_{PSF})$ for different values of σ for the above two cases are shown in Figs. 6(a) and 6(b), respectively. From these results, it is clear that the auto-correlation profile strongly depends on the scattering degree, and each profile is far from being delta-like function, irrespective to the QPM presence.

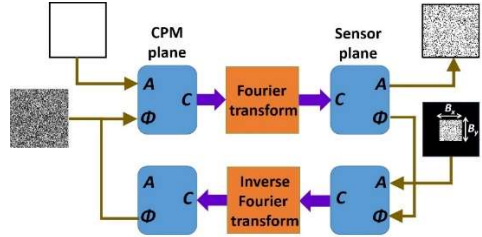

Fig. 2. Modified Gerchberg Saxton algorithm for synthesizing coded phase masks with different scattering degrees.

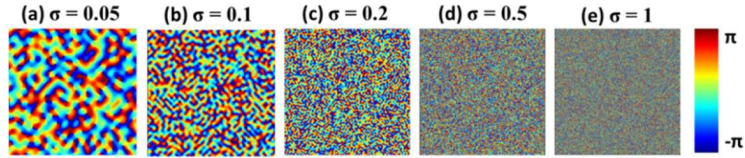

Fig. 3. Phase images of the pseudorandom coded phase masks with different scattering degrees synthesized by GSA.

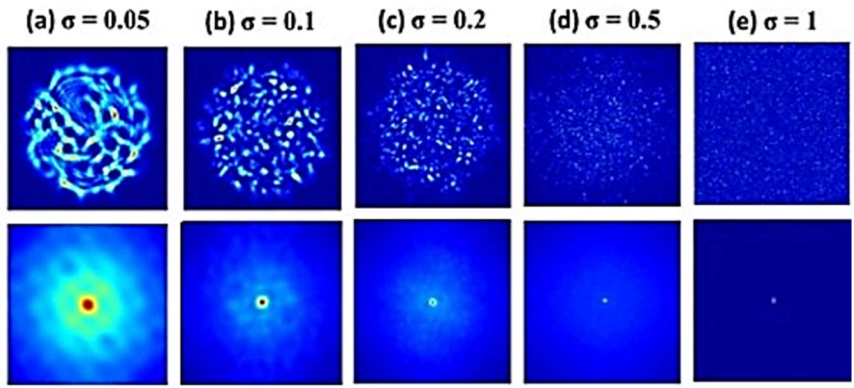

Fig. 4. Images of simulated PSFs (top row) and their respective autocorrelation images (bottom row) when only a single scattering medium is positioned in the system. Every image consists of 500 ×500 pixels.

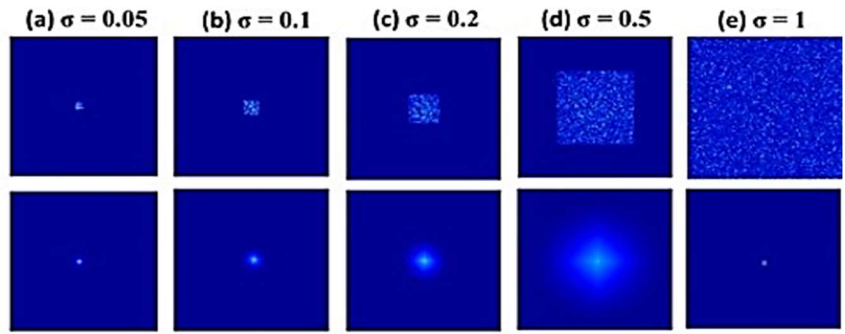

Fig. 5. Images of simulated PSFs (top row) and their respective autocorrelation images (bottom row) when a single scattering medium with attached QPM is positioned in the system. Every image consists of 500 ×500 pixels.

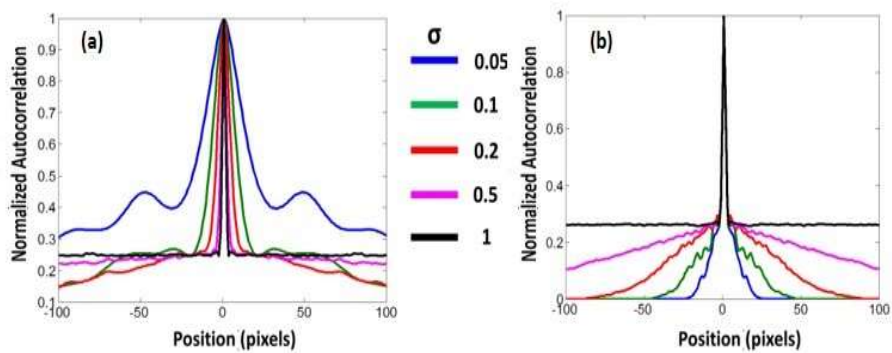

Fig. 6. Plot of the normalized cross-sections of $I_{PSF} \otimes I_{PSF}$ for different values of σ for (a) a single scattering medium of Fig. 4 and (b) a single scattering medium with QPM of Fig. 5.

From Fig. 6(a), it is evident that without QPM, as much as the scattering degree has increased the autocorrelation is narrower. In both cases of Figs. 6(a) and 6(b), the central peak is accompanied by a background, which becomes more uniform with the increment of the scattering degree. Therefore, to guarantee an appropriate convergence of the phase retrieval algorithm, one should eliminate the background. Following previous methods of eliminating the background [14], we display on the SLM two independent CPMs and subtract one intensity response from the other. The result is a bipolar PSF that its autocorrelation is close to a delta function with a negligible background, as is shown in Fig. 7.

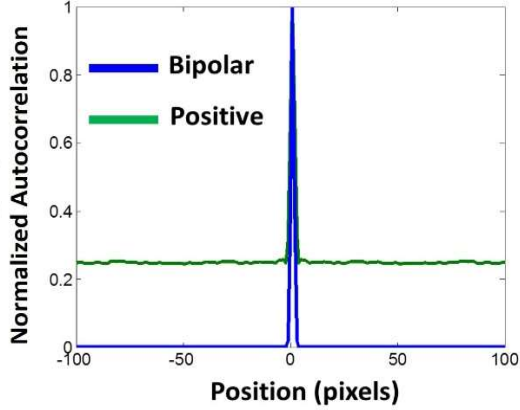

Fig. 7. Autocorrelation profile of positive PSF (green) and bipolar PSF (blue).

From the above study, we conclude that the autocorrelation profile can be engineered using the proposed technique. In the following section, the phase retrieval algorithm is discussed.

*2.3 Phase retrieval algorithm*

The autocorrelation of the object intensity pattern on the image sensor in the presence of a single scattering layer, with or without QPM, can be expressed as [21],

$$\begin{aligned} I_{OBJ} \otimes I_{OBJ} &= \left(I_{PSF} * o(\bar{r})\right) \otimes \left(I_{PSF} * o(\bar{r})\right) \\ &= \left(I_{PSF} \otimes I_{PSF}\right) * \left(o(\bar{r}) \otimes o(\bar{r})\right) \\ &\cong \left(o(\bar{r}) \otimes o(\bar{r})\right) + B(\bar{r}), \end{aligned} \quad (7)$$

where $B(\bar{r})$ is the background function. The above approximation is valid only when $\left(I_{PSF} \otimes I_{PSF}\right)$ yields a delta-like function plus a background. In the proposed technique, the CPM can be used to engineer the autocorrelation profile of the PSF to be as sharp as possible and the background $B(\bar{r})$ can be almost removed by the use of a bipolar intensity pattern as shown in Fig. 7. Based on Eq. (6) the bipolar PSF is,

$$\begin{aligned} I_{PSF}(\bar{r}_0) &= I_{PSF,1}(\bar{r}_0) - I_{PSF,2}(\bar{r}_0) \\ &= \left| \mathfrak{F}\left\{ D\left(\frac{\lambda z_h z_s \bar{r}}{u}\right) \cdot C_1(\lambda z_h \bar{r}) \right\} \right|^2 - \left| \mathfrak{F}\left\{ D\left(\frac{\lambda z_h z_s \bar{r}}{u}\right) \cdot C_2(\lambda z_h \bar{r}) \right\} \right|^2. \end{aligned} \quad (8)$$

Thus, the autocorrelation of the scattered light can be represented as:

$$\begin{aligned}
H_{OBJ} \otimes H_{OBJ} &= (I_{OBJ,1} - I_{OBJ,2}) \otimes (I_{OBJ,1} - I_{OBJ,2}) \\
&= I_{OBJ,1} \otimes I_{OBJ,1} - I_{OBJ,1} \otimes I_{OBJ,2} - I_{OBJ,2} \otimes I_{OBJ,1} + I_{OBJ,2} \otimes I_{OBJ,2} \\
&= [I_{PSF,1} * o(\bar{r})] \otimes [I_{PSF,1} * o(\bar{r})] - 2B(\bar{r}) + [I_{PSF,2} * o(\bar{r})] \otimes [I_{PSF,2} * o(\bar{r})] \\
&= [I_{PSF,1} \otimes I_{PSF,1}] * [o(\bar{r}) \otimes o(\bar{r})] - 2B(\bar{r}) + [I_{PSF,2} \otimes I_{PSF,2}] * [o(\bar{r}) \otimes o(\bar{r})] \\
&\cong a_1 \delta(\bar{r}) * [o(\bar{r}) \otimes o(\bar{r})] + B(\bar{r}) - 2B(\bar{r}) + a_2 \delta(\bar{r}) * [o(\bar{r}) \otimes o(\bar{r})] + B(\bar{r}) \\
&\cong a[o(\bar{r}) \otimes o(\bar{r})] \cong a \mathfrak{F}^{-1}\left\{\left|\mathfrak{F}\{o(\bar{r})\}\right|^2\right\}
\end{aligned} \quad (9)$$

where $\mathfrak{F}^{-1}$ indicates the inverse 2D Fourier transform and $a_1$, $a_2$, $a$ are constants. In Eq. (9), it is assumed that regardless of the precise CPM distribution, as long as the statistical parameters are the same for the two CPMS, the background function $B(\bar{r})$ is approximately the same for the two autocorrelations of the object intensities and for the cross-correlation between them. Based on Eq. (9) the magnitude of the object's spectrum is approximately,

$$\left|\mathfrak{F}\{o(\bar{r})\}\right| \cong a^{-1}\sqrt{\mathfrak{F}\{H_{OBJ} \otimes H_{OBJ}\}} \quad . \quad (10)$$

The spectral magnitude of the object is fed into the phase retrieval algorithm [27], where the missing spectral phase is evaluated by this iterative algorithm. The size of the object is approximately half of the size of the autocorrelation pattern given in Eq. (9) divided by the system magnification $M_T$. Hence, by measuring the size of the autocorrelation pattern $2M_T \cdot (w_x, w_y)$, one can estimate the limiting window of the size $(w_x, w_y)$ on the object plane of the iterative algorithm.

The phase retrieval algorithm shown in Fig. 8 begins with an initial random magnitude matrix over the limited window and zero values over the rest of the object plane. The object plane information is Fourier transformed and the magnitude is replaced by $\sqrt{\mathfrak{F}\{H_{OBJ} \otimes H_{OBJ}\}}$, while the phase information is retained. The modified complex amplitude is inverse Fourier transformed and non-negative and real values constraints are applied [20]. This process is repeated until hopefully an image similar to the object is retrieved.

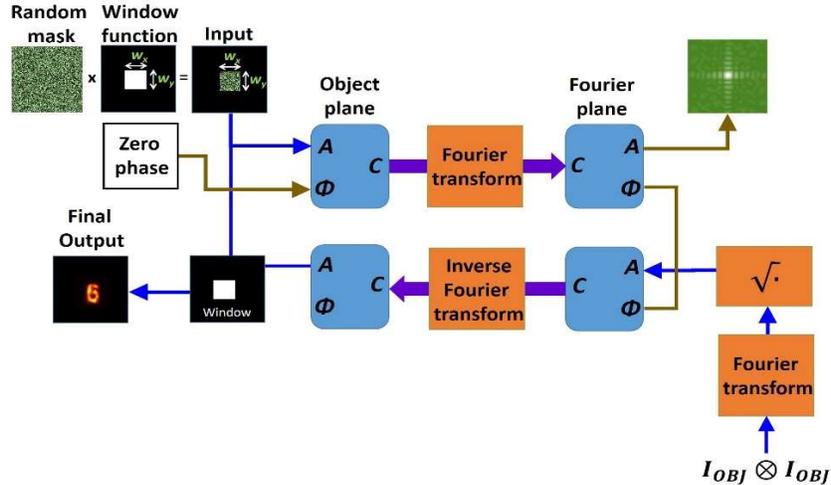

Fig. 8. Modified phase retrieval algorithm for retrieving the image of the hidden object.

## 3. Experiments

The experimental setup for the proposed method is shown in Fig. 9. Light emitted from a solid-state laser ($\lambda = 532\ nm$) is passed through a rotating diffuser. From the diffuser, it was collected by a refractive lens $L_1$ to critically illuminate the object. Initially, we had used a pinhole with a diameter of 100 $\mu m$ in the object plane to validate our assumptions and later, we used a more complex object. The light diffracted by the object passed through a constant ground glass diffuser with 1500 grit (Thorlabs) located at a distance of about 1 $mm$ from the object. The diffused light was projected on a phase-only reflective SLM [Jasper EDucation Kit (EDK) – JD955B, 1920×1080 pixels, 6.54 $\mu m$ pixel pitch] by a relay system comprising of two identical refractive lenses $L_2$ and $L_3$ with a diameter of 5 $cm$ and focal length of $f$=10 $cm$. The light emitted from the diffuser further propagated a distance of $z_r$=16 $cm$ before reaching the relay system, which projected the light onto the SLM plane. A polarizer $P$ placed beyond $L_3$ was used to polarize the light along the orientation of the SLM active axis to maximize the modulation efficiency. A quadratic phase mask with a focal length of 6.9 $cm$ was multiplied with CPM synthesized using GSA and both were displayed on the SLM. The reflected light from the SLM was recorded by an image sensor (GigE vision GT Prosilica, 2750×2200 pixels, 4.54 $\mu m$ pixel pitch) located at a distance of $z_h$=11.7$cm$, from the SLM.

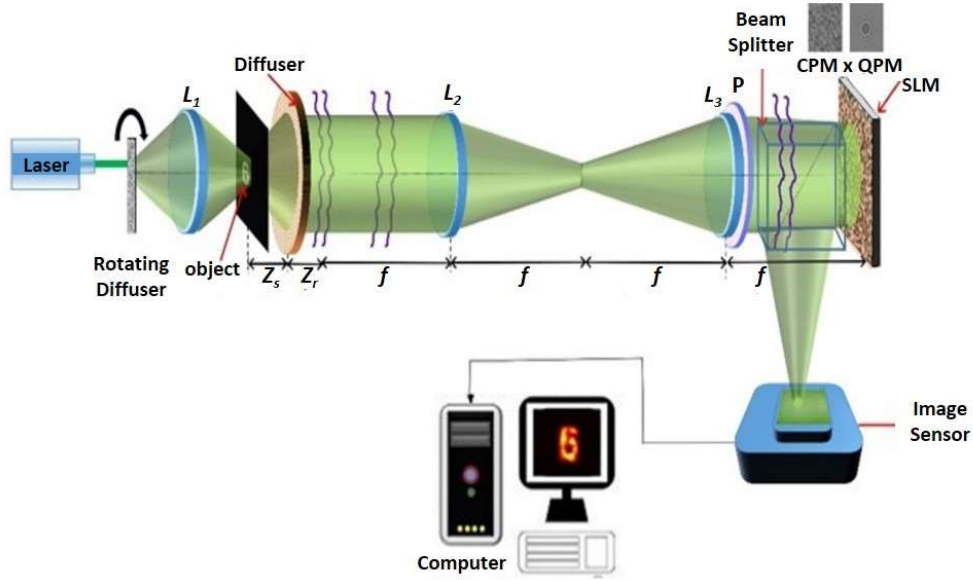

Fig. 9. Experimental setup; CPM – Coded phase mask; QPM – Quadratic phase mask; $L_1$, $L_2$, $L_3$ – Refractive lenses; $P$ – Polarizer; SLM – Spatial light modulator.

In the first experiment with the pinhole of 100 μm diameter, three cases were studied. In case 1, only a quadratic phase mask was displayed on the SLM and the intensity pattern ($I_1$) at the sensor plane was recorded. In case 2, CPM with a scattering degree of σ= 0.2 was displayed with the QPM and the intensity pattern ($I_2$) was recorded. In case 3, a second independent CPM was combined with the QPM and another intensity pattern ($I_{2b}$) was recorded. A bipolar pattern ($I_3 = I_2 - I_{2b}$) was calculated and stored. The patterns and their autocorrelation for the three cases are shown in Fig. 10. It is clearly seen that the autocorrelation result of the bipolar pattern in Fig. 10(c) has the closest resemblance to the autocorrelation of the pinhole image shown in Fig. 10(d).

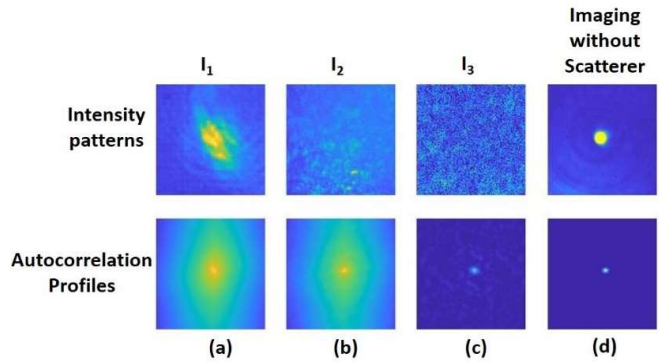

Fig. 10. Images of intensity patterns and their respective autocorrelations, (a) $I_1$ denotes intensity pattern recorded by the sensor when only QPM is displayed on the SLM, (b) $I_2$ denotes intensity pattern recorded by the sensor when QPM and CPM are displayed on the SLM, (c) $I_3$ denotes a bipolar pattern obtained by $I_2 - I_{2b}$, and (d) direct imaging of the pinhole without any physical diffuser. Every image consists of 500 ×500 pixels.

In the next experiment, the digits '2' and '6' from group 4 of United States Airforce (USAF) resolution target were tested. The objects were placed one by one at the object plane behind the static diffuser. First, only QPM was displayed on the SLM and the intensity pattern shown in Figs. 11(a) and 11(d) was recorded by the sensor. In these figures, the digits are not recognizable and hence a method of image recovery is required. Next, the phase retrieval algorithm was applied first without CPM and later with a single CPM, in order to convince that the dual CPM method is indeed desirable. For the case of QPM only, the autocorrelation of the recorded intensity pattern was Fourier transformed and its square root was fed into the phase retrieval algorithm shown in Fig. 8. The phase retrieval algorithm ran for 400 iterations, the number of iterations which in the later experiment with the two CPMs guarantees convergence. A random initial matrix over the limited window and zero values over the rest of the object plane is fed into the algorithm. The results for 25 different random initial matrices are shown in Figs. 11(b) and 11(e). The results of Figs. 11(b) and 11(e) are obtained from the first 25 experiments without omitting any result. In the current experiment with QPM only, shown in Figs. 11(b) and 11(e), it is clearly seen that none of the reconstructed results are close to the direct image of the objects without the diffuser as shown in Figs. 11(c) and 11(f) with a magnification of 0.74.

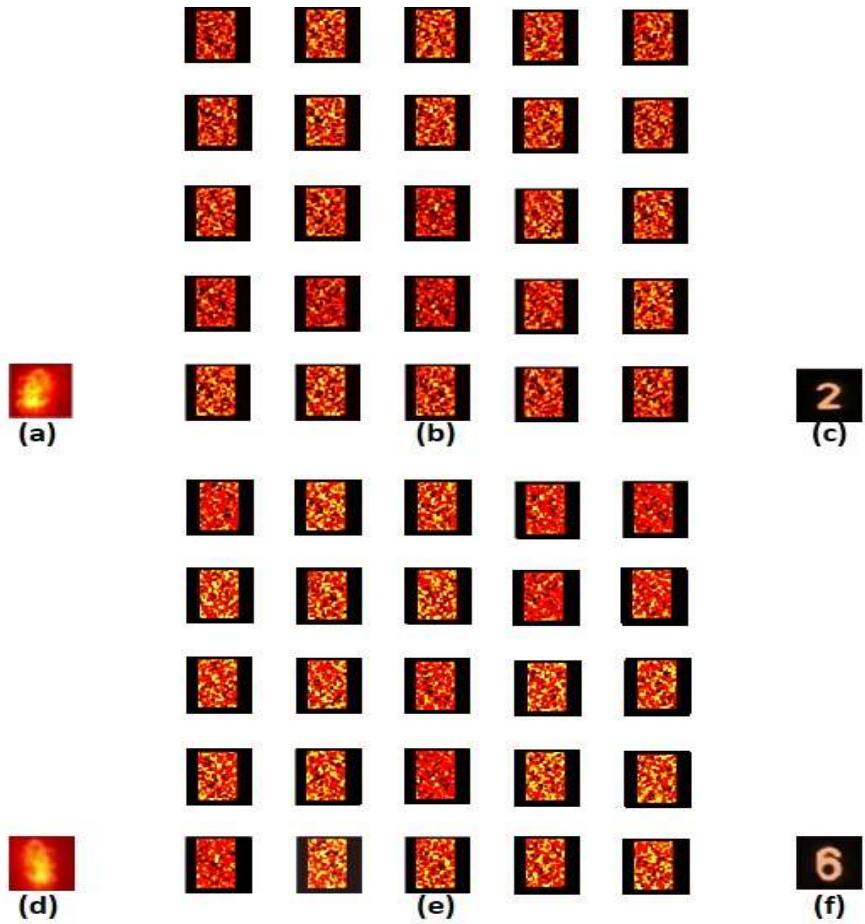

Fig. 11. (a) and (d): Intensity pattern on the sensor plane with the diffuser and only QPM on the SLM for the digit '2' and for the digit '6', respectively, (b) and (e) : reconstruction results with the diffuser and only QPM on the SLM with 25 different initial random matrices for the digit '2'and for the digit '6', respectively, (c) and (f) direct image of the object without the diffuser and with QPM displayed on the SLM.

A CPM was synthesized using the GSA, with σ=0.2. This scattering degree yielded the best reconstruction results with our proposed method among the five tested values of σ. In the second test, a single CPM and the QPM were displayed on the SLM and the corresponding intensity pattern was recorded. The light scattered by the diffuser and recorded by the image sensor was modulated by both the CPM and QPM displayed on the SLM. The reconstruction results with 25 initial random matrices are shown in Fig. 12. In this case, a slight improvement in the results can be observed, as compared to the previous case.

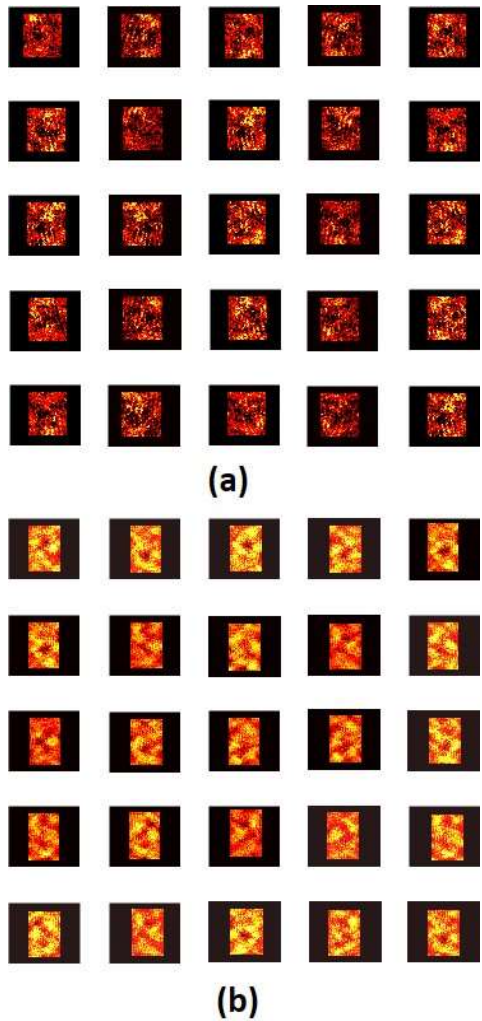

Fig. 12. (a) Reconstruction results obtained from the phase retrieval algorithm with 25 initial random matrices for the digit '2', (b) reconstruction results obtained from the phase retrieval algorithm with 25 initial random matrices for the digit '6'; QPM and single CPM are displayed in the SLM.

The last part of the experiment involves the recording of two intensity patterns corresponding to two different CPMs both with σ = 0.2. One pattern was subtracted from the other to yield the bipolar pattern. The bipolar pattern indeed reduces the bias term from the autocorrelation as is shown in Fig. 13(c) in comparison to the other two cases of Fig. 13(a) without CPM and of Fig. 13(b) with a single CPM. Thus, a significant improvement of the reconstruction results over the other two cases is expected.

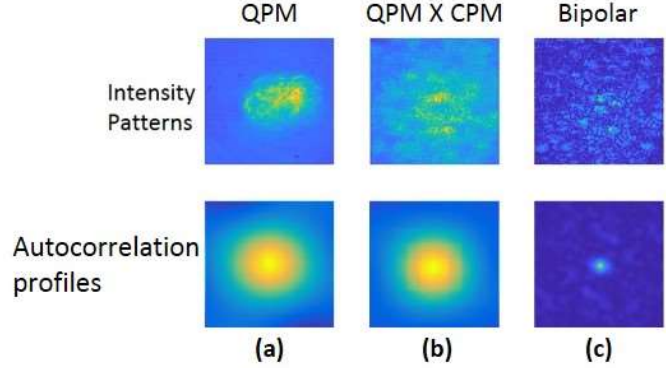

Fig. 13. Images of intensity patterns and their respective autocorrelation profiles. (a) Intensity pattern recorded by the sensor and its autocorrelation pattern when QPM is displayed on the SLM. (b) Intensity pattern recorded by the sensor and its autocorrelation pattern when QPM and CPM are displayed on the SLM. (c) Bipolar pattern and its autocorrelation pattern. Every image consists of 500 ×500 pixels. Intensity patterns correspond to digit 6 from group 4 of United States Airforce (USAF) resolution target.

The same phase retrieval algorithm was applied on the autocorrelation of the bipolar matrix. The reconstruction results for both the objects '2' and '6', with 25 different initial random are shown in Figs. 14(a) and 14(c), respectively. The results of Figs. 14(a) and 14(c) are also obtained from the first 25 experiments without omitting any result. Furthermore, for both groups of images, averaging was done over the complex images to improve the quality of results. In order to average, the center of mass was calculated for every reconstructed image, and each image was shifted such that all the images had a common center of mass. The averaged reconstruction result of all the images is shown in Figs. 14(b) and 14(d).

## 4.  Summary and conclusions

We have proposed and demonstrated a noninvasive technique to image through a thin scattering layer, using a two user-controlled phase coded masks with two camera shots. The explanation to the success of the two-shot method is as following; in each measured intensity pattern there is a dominant bias function, which does not contain any information on the hidden object, but on the other hand, it obscures the relatively small signal, which does contain the object information. The difference between the two measured intensity patterns created from two independent, but statistically similar CPMs, contains only negligible remains from the bias function, because there is not much difference between the entire bias functions. On the other hand, the autocorrelation of the difference between the two measured intensities does contain the information of the covered object autocorrelation as is indicated by Eq. (9), and hence the object itself can be reconstructed by the phase retrieval algorithm.

The method can be easily applied for imaging through different types of scatterers by accurately engineering the autocorrelation profile of the scattered light, using suitable CPMs. There is a higher level of ambiguity present in the reconstruction of the object when a single camera shot is used in the phase retrieval algorithm. The proposed method substantially reduces the ambiguity present in the reconstruction. As a result, the phase retrieval algorithm is not needed to be executed with hundreds of initial random phase masks, each runs several thousand iterations like in [21]. We believe that the proposed technique will be useful for imaging through scattering layers with lesser complexity and time compared to the existing techniques. In this study, the method is only applied for 2D imaging. Therefore, additional studies are required to improve the system performances toward 3D imaging systems.

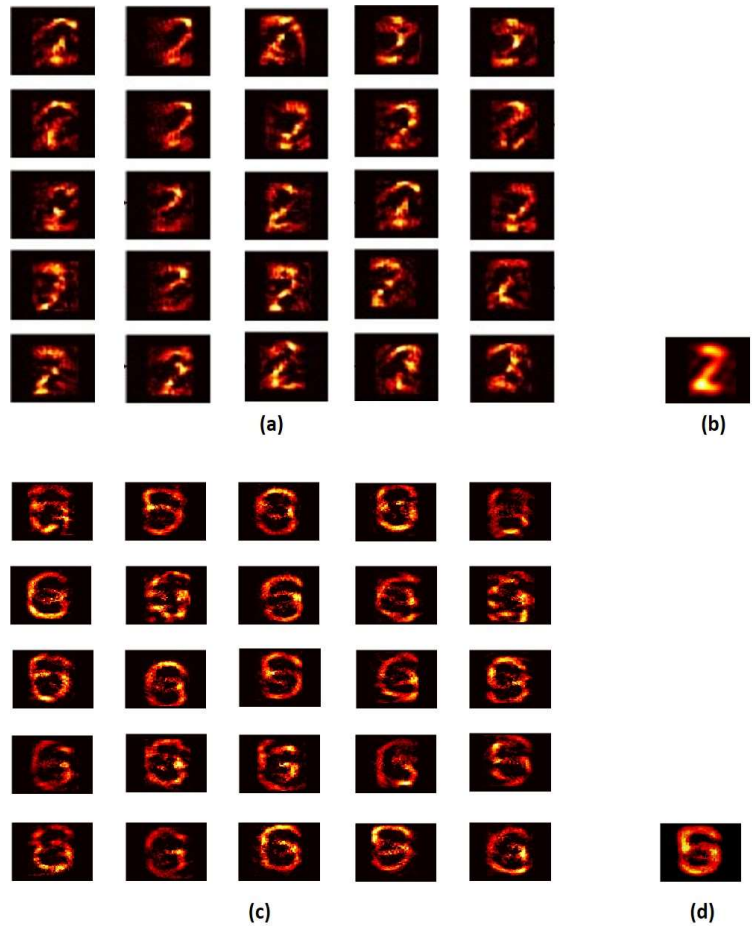

Fig. 14 (a) and (c): Reconstruction results with bipolar intensity patterns using the phase retrieval method (b) and (d): Averaged reconstruction result.

**Funding**

The work was supported by the Israel Science Foundation (ISF) (Grants No. 1669/16); Israel Ministry of Science and Technology (MOST).

**Acknowledgment**

This study was done during a research stay of J.R. at the Alfried Krupp Wissenschaftskolleg Greifswald.